\documentstyle[prl,aps,twocolumn,floats,amssymb,epsfig]{revtex}

\def\bbZ{{\mathbb Z}}

\def\SU{{\rm{SU}}}
\def\U{{\rm{U}}}
\newcommand{\mb}[1]{\ifmmode#1\else\mbox{$#1$}\fi}

\newcommand\al{\mb{\alpha}}

\newcommand\be{\mb{\beta}}
\newcommand\ga{\mb{\gamma}}
\newcommand\de{\mb{\delta}}

\newcommand\la{\mb{\lambda}}

\newcommand\si{\mb{\sigma}}

\newcommand\om{\mb{\omega}}

\newcommand\La{\mb{\Lambda}}

\newcommand\Om{\mb{\Omega}}

\newcommand\calC{\mb{{\cal C}}}

\newcommand\calH{\mb{{\cal H}}}

\newcommand\calL{\mb{{\cal L}}}

\newcommand\calO{\mb{{\cal O}}}
\newcommand\calP{\mb{{\cal P}}}

\newcommand\calZ{\mb{{\cal Z}}}
                    
\newcommand{\beq}{\begin{equation}}
\newcommand{\eeq}{\end{equation}}
\newcommand{\nn}{\nonumber}
\newcommand{\bea}{\begin{eqnarray}}
\newcommand{\eea}{\end{eqnarray}}

\newcommand{\tr}{\mb{{\rm tr}}}

\newcommand{\pderiv}[2]{\frac{\partial {#1}}{\partial {#2}}}
\newcommand{\x}{\mb{\times}}
\newcommand{\rhat}{\hat\bm r}

\newcommand{\Ad}{{\rm Ad}}
\newcommand{\gsim}
{\raise.3ex\hbox{$\;>$\kern-.75em\lower1ex\hbox{$\sim$}$\:$}}
\newcommand{\lsim}
{\raise.3ex\hbox{$\;<$\kern-.75em\lower1ex\hbox{$\sim$}$\:$}}
\newcommand{\ts}{\textstyle}
\newcommand{\half}{\mb{\ts \frac{1}{2}}}
\newcommand{\third}{\mb{\ts \frac{1}{3}}}
\newcommand{\twothird}{\mb{\ts \frac{2}{3}}}

\newcommand{\fifth}{\mb{\ts \frac{1}{5}}}
\newcommand{\fourfifth}{\mb{\ts \frac{4}{5}}}

\newcommand{\bm}[1]{{\mbox{\boldmath $#1$}}}
\newcommand{\sm}{{\rm SM}}
\newcommand{\rmC}{{\rm C}}
\newcommand{\rmI}{{\rm I}}
\newcommand{\rmY}{{\rm Y}}
\newcommand{\minus}{{\mbox{$$-$$}}}

\begin{document}
\draft


\twocolumn[\hsize\textwidth\columnwidth\hsize\csname @twocolumnfalse\endcsname
\title{Monopoles and Dyons in SU(5) Gauge Unification\\
(with relation to a Dual Standard Model)}
\author{Nathan\  F.\  Lepora\footnotemark}
\address{101 Larkhall Rise, Clapham, England}
\date{February 7, 2001}
\maketitle

\begin{abstract}
This article presents a fairly complete description of the monopoles and 
dyons arising from an SU(5) gauge unification of the standard model. Topics
discussed include: the spectrum of SU(5) monopoles; their gauge equivalence 
structure; their spherical symmetry; the construction of dyons; global 
charge; the intrinsic angular momenta of scalar boson-monopole composites
and monopole gauge excitations; and the effects of a theta vacuum. The 
relevance of each of these topics to constructing a dual description
of the standard model with SU(5) solitons is discussed in detail.
\end{abstract}
\pacs{pacs no.s.}]



\section{Introduction}
\footnotetext[1]{\ email: n$\_$lepora@hotmail.com}

The concept of magnetic charge has presumably been around since that
of magnetic field. However it was Dirac who discovered the magnetic 
{\em monopole}~\cite{dirac}. He showed that the gauge configuration 
$A_\varphi\sim m(1-\cos\theta)$
describes an isolated magnetic charge $m$. Implicit in this description is a 
string singularity, which Dirac famously showed to be unobservable
when the magnetic charge is a multiple of $2\pi/e$.

Whilst this motivated much interest, monopoles didn't achieve their
present status until 't~Hooft and Polyakov's celebrated work~\cite{hooft}. 
They demonstrated such monopoles occur as solitons in an SU(2) gauge 
theory spontaneously broken to U(1). As particle physics is based on similar
theories, monopoles are thus immediately relevant.

At present monopoles are mainly used in two applications: 
confinement and gauge unification. This article will concentrate
on the latter of these, where monopoles are an inevitable consequence of  
unifying the fundamental gauge interactions~\cite{kib}. 

Since 't~Hooft and Polyakov's work such monopoles have taken
a wider significance within field theory. In particular their behaviour
has a quite unparalled richness of structure. For instance their effects
include: the spin from
isospin mechanism, leading to spin half configurations in a bosonic gauge
theory~\cite{jackiw:spin,gold:spin}; the problem of global charge, whereby 
some charges are not properly defined around a
monopole~\cite{globcol,bala}; and the effects of a theta vacuum, which 
converts pure monopoles into dyons~\cite{witten}. 

Another important property of monopoles is their electric-magnetic duality,
where a system of electric or magnetic charges behave identically. 
Consequently monopoles offer an alternative description of particles.
That is, a charged particle is usually considered to  
source $A^0\sim e/4\pi r$, but instead the dual gauge potential
(such that $\bm E = \bm\nabla\wedge\tilde\bm A$) can be used
to represent it as a monopole $\tilde A_\varphi\sim e(1-\cos\theta)/4\pi$. 

Whilst electric-magnetic duality is fairly established within Abelian theories,
there are unsolved issues in the non-Abelian case; although certainly
such a duality should exist. Recently, however, Chan and Tsou have described 
an exact non-Abelian electric-magnetic duality~\cite{chandual}.
They conclude the magnetic gauge symmetry is the same group as the 
electric gauge symmetry, although its gauge potential has the opposite parity.

The existence of this non-Abelian duality
has a wider significance for particle physics, since it offers an 
alternative method for formulating the standard model. Such a theory would
represent the properties of elementary particles by the behaviour
of monopoles. This theory, as proposed by Liu and Vachaspati, would be a 
{\em dual standard model}~\cite{vachdsm}. 

It is quite possible that the construction of a dual standard model may
uncover a hidden simplicity and regularity of form that underlies
the conventional standard model. Such a hidden structure could prove 
crucial to understanding the nature and origin of the elementary particles. 
Also possible is that new physics may have to be 
included to arrive at a simple and consistent form.

A primary indication of the dual standard model's structure is from a 
remarkable discovery by Vachaspati. He observed that the magnetic charges 
of the
five stable monopoles within Georgi-Glashow SU(5) unification are identical
to the electric charges in one generation of elementary 
particles~\cite{vach95}. That is, a 
unification of the standard model's magnetic gauge sector leads to a 
spectrum of electric monopoles whose charges precisely mimic 
one generation of elementary particles.

This lead Vachaspati to propose that perhaps the dual 
standard model could be based on the properties of SU(5) monopoles. 
In this sense the elementary particles would then originate as solitons from 
magnetic gauge unification. If successful such a theory
would offer a particularly
simple and elegant method for unifying the gauge and matter content of the
standard model.

However there are more aspects to the standard model than just its
charge spectrum; for instance spin, confinement, mixing, particle mass, 
chirality and parity violation. A successful dual standard model must either
incorporate or explain these properties. At present this is an ongoing
task, although many of these features are naturally falling into place.
For instance spin can be included through the spin from isospin 
mechanism~\cite{vachspin}, and confinement has a natural description as a dual 
Meissner effect~\cite{vachdsm,gold99}. In addition some potential explanations 
for other features have been proposed by Vachaspati and Steer~\cite{steer} 
and by the author~\cite{comp}.

To fully investigate these effects one must have a complete understanding
of the monopoles from Georgi-Glashow 
$\SU(5)\rightarrow\SU(3)\x\SU(2)\x\U(1)/\bbZ_6$ gauge unification. 
However there is no review that provides a reasonably complete picture of such 
SU(5) monopoles. Hopefully this article will address that issue, 
whilst also providing some background to the dual standard model. 

In several ways this article relates to Dokos and Tomaras's original 
description of monopoles and dyons arising within 
$\SU(5)\rightarrow\SU(3)\x\U(1)/\bbZ_3$~\cite{dokos}. 
Their work formed the basis for much research
into the implications of monopoles within grand unification. 
As this article relates to a different phenomenological situation it takes
a different slant on the problem. Additionally there have been several 
important results that have occurred since their work was completed.

The composition of this article is as follows. Firstly the monopoles in 
SU(5) gauge unification are considered. Sec.~(\ref{sec:dsm}) describes
their spectrum and similarities to the elementary particles.
Sec.~(\ref{sec:monopole}) discusses their gauge transformation properties, 
finding representations compatible with the elementary particles. 
Sec.~(\ref{sec:sph}) discusses their spatial rotation properties, 
motivating such monopoles are scalars. Then SU(5) dyons are described. These
correspond to two types: scalar boson-monopole composites, discussed in
sec.~(\ref{sec:spin}); and monopole gauge excitations, discussed in 
sec.~(\ref{sec:gspin}). It appears that both types of dyon 
may have intrinsic half-integer angular momenta. Finally, 
sec.~(\ref{sec:theta}) describes the effects of a theta vacuum.

Before starting note that each of the three
classic review articles~\cite{mon} uses a slightly different convention;
this article follows Preskill, in line with the work on the
dual standard model. 

\section{SU(5) Monopoles}
\label{sec:dsm}

To start it is important to understand the spectrum of stable monopoles 
from SU(5) gauge
unification. The complete spectrum was first obtained by Gardner and
Harvey~\cite{gard84}, who showed there are precisely five stable
monopoles. More recently it has observed that these have magnetic
charges in coincidence with the electric charges in one generation of
elementary particles~\cite{vach95}. This motivates
that perhaps the elementary particles originate as monopoles
from a magnetic gauge unification of the standard model.

The Georgi-Glashow SU(5) gauge unification can be described through the 
following symmetry breaking~\cite{gg}
\beq
\label{dualsb}
\SU(5) \stackrel{24}{\longrightarrow} H_{\rm SM}
=[\SU(3)_{\rm C} \x \SU(2)_{\rm I} \x \U(1)_{\rm Y}]/\bbZ_6.
\eeq
This symmetry breaking is achieved through condensation of an adjoint
scalar field $\Phi$; then with respect to the vacuum
$\Phi_0 = iv\, {\rm diag}(\third, \third, \third, -\half, -\half)$ the
standard model gauge symmetry is contained within SU(5) as   
\beq
\label{Hembed}
\left( 
\begin{array}{cc} 
\SU(3)_{\rm C} & 0 \\ 0 &  \SU(2)_{\rm I} 
\end{array} 
\right) \x \U(1)_Y,
\eeq
with $\U(1)_Y$ along the diagonal. In the breaking
(\ref{dualsb}) a feature to bear in mind is the discrete  
$\bbZ_6$ quotient: this represents an intersection between the
colour-isospin and the hypercharge parts of (\ref{Hembed}). Then, since 
$\bbZ_6$ is included twice in (\ref{Hembed}) but only once in SU(5), it
divides out in (\ref{dualsb}). 

That monopoles occur is implied by the non-trivial topology 
of (\ref{dualsb}), where each distinct monopole corresponds to a second 
homotopy class
\bea
\label{homotopy}
\pi_2 \left( \frac{\SU(5)}{H_{\rm SM}} \right) 
&\cong& \pi_1(H_{\rm SM}).
\eea
Within this topology the $\bbZ_6$ quotient in (\ref{dualsb})
determines the basic pattern of monopoles.

To find the monopole spectrum it is convenient to 
associate each monopole with a magnetic generator $M$ 
\beq
\label{B_top}
\Phi \sim \Phi_0,\ \ \ \ {\bm B} \sim \frac{1}{2g} \frac{\rhat}{r^2} M.
\eeq
This is in a unitary gauge, so there is an implicit Dirac string
in the gauge potential. To have a well-defined solution this Dirac string must
be a gauge artifact, which constrains $M$ through a topological 
quantisation~\cite{engl76}  
\beq
\label{qq}
\exp(i2\pi M) =1.
\eeq
As such $M$ has integer eigenvalues. Additionally, a finite energy monopole
has a massless long range magnetic field; hence $M$ is a generator of
$H_{\rm SM}$. 

The individual colour, isospin and hypercharge magnetic charges are defined
by a gauge choice that the monopole's magnetic field takes the form 
\beq
{\bm B} = T_\rmC {\bm B}_\rmC + T_{\rmC'} {\bm B}_{\rmC'} + 
T_\rmI {\bm B}_\rmI + T_\rmY {\bm B}_\rmY,
\eeq
with generators to be defined below. Then the magnetic charges are
\beq
\label{Q}
M= m_{\rm C}T_{\rm C} + m_{\rm C'}T_{\rm C'} + 
m_{\rm I}T_{\rm I} + m_{\rm Y}T_{\rm Y}. 
\eeq
Within this definition care must be taken with the normalisation of each
$T$. To ease later sections of this review a normalisation $\tr\,{T^2}=1$ is
taken, which is slightly different from that used in
similar papers. In that case a suitable choice of generators
is 
\bea
T_{\rm C}&=& \surd\ts\frac{3}{2}\, \la_8 
= \surd\ts\frac{3}{2} \,{\rm diag}
(\minus\third, \minus\third, \twothird, 0,0),\nn\\ 
T_{\rm C'}&=& \surd\ts\frac{1}{2}\, \la_3 
= \surd\ts\frac{1}{2} \,{\rm diag}(1, \minus 1,0,0,0),\nn\\ 
\label{gen}
T_{\rm I}&=& \surd\ts\frac{1}{2}\, \si_3 = 
\surd\ts\frac{1}{2}\,{\rm diag}(0,0,0, \minus 1, 1),\\
T_{\rm Y}&=& \surd\ts\frac{5}{6}\, \si_0 = 
\surd\ts\frac{2}{15} \,{\rm diag}
(1,1,1,\minus\ts\frac{3}{2},\minus\ts\frac{3}{2},),\nn
\eea
which for convenience are taken to be diagonal and consistent with
(\ref{Hembed}).

The calculation of $M$ for each monopole then
becomes a determination of which sets  
$(m_{\rm C},m_{\rm C'}, m_{\rm I}, m_{\rm Y})$ solve (\ref{qq}); this leads
straightforwardly to the following magnetic charges for the first six homotopy
classes of $\pi_1(H_{\rm SM})$ \cite{gard84}:

\begin{table}[h]
\caption{Monopole charges and their associated elementary particles.}
\begin{eqnarray*}
\begin{array}{|c|ccc|c|ccc|c|}
\hline
\pi_1 &m_{\rm C}&m_{\rm I}&m_{\rm Y}& {\rm diag}\  M &
n_{\rm C}&n_{\rm I}&n_{\rm Y}& \\
\hline
\vspace*{-0.35cm} &&&&&&&& \\
1 & \surd\frac{2}{3} & \surd\frac{1}{2} & \frac{1}{3}\surd\frac{15}{2} &
(0,0,1,$-$1,0) & 1 & \half & \third &\  (u, d)\  \\
\vspace*{-0.35cm} &&&&&&&& \\
2 & $-$\surd\frac{2}{3} & 0 & \frac{2}{3}\surd\frac{15}{2} &
(0,1,1,$-$1,$-$1) &$-1$ & 0 & \twothird & \bar{d} \\
\vspace*{-0.35cm} &&&&&&&& \\
3 & 0 & $-$\surd\frac{1}{2} & 1\surd\frac{15}{2} &
(1,1,1,$-$2,$-$1) &0 & $-$\half & 1 & (\bar{\nu}, \bar{e}) \\
\vspace*{-0.35cm} &&&&&&&& \\
4 & \surd\frac{2}{3} & 0 & \frac{4}{3}\surd\frac{15}{2} &
(1,1,2,$-$2,$-$2) &1 & 0 & \ts\frac{4}{3} & u \\
\vspace*{-0.35cm} &&&&&&&& \\
5 & - & - & - & - & - & - & - & - \\
\vspace*{-0.35cm} &&&&&&&& \\
6 & 0 & 0 & 2\surd\frac{15}{2} &
(2,2,2,$-$3,$-$3) & 0 & 0 & 2 & \bar{e} \\
\hline 
\end{array}
\end{eqnarray*}
\vspace*{-0.5cm}
\label{tab1}
\end{table}

\noindent
A few remarks are in order about these monopoles:

\noindent (i)
To simplify the comparison with the elementary particles their
charges are expressed in a basis with simpler normalisation; having 
$n_\rmC=\surd\frac{3}{2}\,m_\rmC$, $n_\rmI=\surd{2}\,m_\rmC$ and 
$n_\rmY=\surd\frac{2}{15}\,m_\rmY$. These normalisations play a central role in
an associated gauge unification~\cite{meunif}.

\noindent (ii) 
Gardner and Harvey have shown that only the above five monopoles are stable
for a wide parameter range~\cite{gard84}; with the 5 and the
$n\geq 7$ monopoles unstable to fragmentation. Further non-topological
charge may be added to the above monopoles, for instance taking  
$m'_{\rm C}=m_{\rm C}+3$ or $m'_{\rm I}=m_{\rm I}+2$, but by Brandt and Neri's
stability analysis these are unstable to long range gauge 
perturbations~\cite{brandt}.

\noindent (iii) 
Splitting $\bbZ_6$ into colour and isospin factors $\bbZ_3 \x \bbZ_2$
exhibits the connection between the topology and the monopole spectrum: 
giving a $1,-1,0,\cdots$ periodicity of $n_{\rm C}$ and the 
$\half,0,\cdots$ periodicity of $n_{\rm I}$.

\noindent (iv)
For a monopole with
non-zero colour there are in fact a triplet of $(m_{\rmC'},m_\rmC)$ colour
charges $(0,\surd\frac{2}{3})$ and
$(\pm\surd\frac{1}{2},-\surd\frac{1}{3})$. In addition each monopole with 
non-zero isospin is a member of a doublet
$m_\rmI=\pm\surd\frac{1}{2}$. All these values are completely in line with
them forming colour and isospin gauge multiplets.

Table~\ref{tab1} is the central result of this section. From it Vachaspati
made the following key observation: {\em The magnetic
charges of the five stable SU(5) monopoles are in complete accordance with
the electric charges of the five quark and lepton multiplets}. To make the
correspondence explicit each monopole is labeled 
by a particle multiplet in the lightest
generation. The spectrum is completed by identifying the
anti-particles with anti-monopoles. 

This observation is remarkable. It is difficult to believe that it
is just a coincidence; the charges identify exactly and by some miracle all
monopoles not in correspondence are unstable. This suggests a deep
connection between the non-perturbative features of the grand unified theory
and the elementary particle spectrum of the standard model. As Vachaspati
conjectured: {\em This correspondence suggests that perhaps grand unification
should be based on a magnetic SU(5) symmetry group with only a bosonic
sector and the presently observed fermions are really the monopoles of that
theory.}

\section{Gauge Freedom}
\label{sec:monopole}

It has long been believed that there is a duality between non-Abelian
electric particles and magnetic monopoles~\cite{gno}.
This is of primary importance to the particle-monopole correspondence 
described in 
sec.~(\ref{sec:dsm}). There the magnetic charges of SU(5) monopoles identify
with the electric charges of the elementary particles, which suggests  
their interactions should also be the same. That is, their interactions appear
to be dual. 

Recently, Chan and Tsou have discovered an exact non-Abelian electric-magnetic
duality transformation~\cite{chandual}. 
Within this they found that the magnetic gauge symmetry has
the same group structure as the electric sector, though the gauge field is
of opposite parity. This supports that the SU(5) monopoles in table~\ref{tab1}
interact under the standard model gauge symmetry $H_\sm$ with 
the associated charges.

This section examines the monopole's gauge freedom under the $H_\sm$
gauge symmetry and compares it to the gauge freedom of the elementary particles
under the same gauge group. By constructing the relevant orbits, the global 
gauge freedom is shown to be precisely the same. It should be noted that
the following arguments represent an improved version of those in 
ref.~\cite{me:gauge}. 

\subsection{Gauge Freedom of the $(u,d)$ Monopole}
\label{sec:orbit}

A simple method for demonstrating that the $(u,d)$ monopole and particle 
multiplet have the same $H_\sm$ gauge freedom is to examine their 
{\em gauge orbits}. These orbits
consist of a collection of states that are rigidly gauge equivalent to one
another, so their geometry is characteristic of the gauge freedom.  
The particle-monopole correspondence is demonstrated upon showing their
gauge orbits are the same. 

To illustrate the concept of a gauge orbit consider firstly the $(u,d)$ gauge
multiplet, which is a tensor product of a colour $\bm 3$ triplet and
a weak isospin ${\bf 2}$ doublet with a hypercharge phase. Its gauge
orbit is generated by acting $H_{\rm SM}$ upon a typical value, say
$q_{ij}=\de_{i1}\de_{j1}$, which gives
\beq
\label{mq0}
\calO_{(u,d)} = H_{\rm SM} \cdot q \cong \frac{H_{\rm SM}}{C(q)},
\eeq
where $C(q)$ leaves $q$ invariant
\beq
\label{c1}
C(q)= \SU(2)_C\x \U(1)_{{\rm Y}-{\rm I}} \x \U(1)_{{\rm I}+{\rm Y}-2{\rm
C}}/\bbZ_2.
\eeq
The appropriate embedding of $C(q)$ is indicated by its colour, isospin and
hypercharge subscripts. 

Now the task is to describe the $(u,d)$ monopole's gauge freedom. Just as
with the $(u,d)$ particle multiplet this freedom is determined by the action
of a set of $H_\sm$ rigid gauge transformations, which collectively generate 
the monopole's gauge orbit.  

For describing the $(u,d)$ monopole's gauge orbit it will be convenient to
express the monopole in a gauge free of the Dirac string. Such a
gauge is the radial gauge, where the asymptotics are simply those of an
SU(2) 't~Hooft-Polyakov Ansatz embedded within SU(5) 
\beq
\label{hoof}
\Phi(\bm r) \sim \Ad[\Om(\rhat)]\Phi_0,\hspace{2em}
\bm B (\bm r) \sim \frac{1}{2g} \frac{\rhat}{r^2}\,\rhat \cdot \vec T.
\eeq
Here $\Om(\rhat)=e^{-i\varphi T_3/2}e^{i\vartheta T_2/2}e^{-i\varphi T_3/2}$
describes the angular behaviour, which is specified by a set of su(2) Pauli
matrices embedded in su(5)  
\beq
\label{emb}
\vec{T} = 
\left( \begin{array}{ccc} 0_2 && \\ & \vec \si &
\\ && 0 \end{array} \right).
\eeq
For consistency with (\ref{B_top}) in the unitary gauge the magnetic
generator $M$ equals $T_3$. 

Because both the scalar and magnetic field are in the adjoint representation
the action of $H_\sm$ has a rather simple form upon (\ref{hoof}):
simply taking $\vec T \mapsto \Ad(h)\,\vec T$. 
This can be interpreted as rigidly moving the monopole through a set of
gauge equivalent embeddings. The collection of these form the gauge orbit  
\beq
\label{M}
\calO_1 \cong \frac{H_{\rm SM}}{C(\vec T)},
\eeq
where $C(\vec T)$ is the subgroup that leaves all three generators $T_i$
invariant
\beq
C(\vec T)= \SU(2)_C\x \U(1)_{{\rm Y}-{\rm I}} \x \U(1)_{{\rm I}+{\rm Y}-2{\rm
C}}/\bbZ_2.
\eeq

Clearly this is the same as (\ref{c1}) for the $(u,d)$ gauge multiplet. 
Therefore the
gauge multiplet and monopole have a compatible gauge freedom under $H_\sm$.

\subsection{Gauge Freedom of the Other Monopoles}
\label{sec:gp}

The analysis of the $(u,d)$ monopole's $H_\sm$ gauge freedom was fairly simple
because it is essentially an SU(2) 't~Hooft-Polyakov monopole embedded in
SU(5). Unfortunately this is not the case for the other monopoles, which
complicates the analogous calculation. 

It is interesting to note, however, that the other SU(5) monopoles are
uncharged under either colour, isospin or both. This suggests that each
monopole can be approximated within an effective symmetry breaking that
includes only those symmetries relevant to their charges. Within these 
effective
theories it then seems reasonable to use the approach of the $(u,d)$ monopole. 

In some sense this assumes that there is a subset of the gauge freedom
that is relevant to the long range magnetic monopole interactions.
Although this approach has not been rigorously justified it does seem
reasonable. Also, crucially, it yields the desired correspondence between
monopoles and gauge multiplets. 

\subsubsection{Gauge Freedom of the $u$ and $\bar d$ Monopoles}

Both the $u$ and $\bar d$ multiplets form colour ${\bf 3}$ triplets,
with some hypercharge. Analogously to (\ref{M}) their gauge orbits are
therefore 
\beq
\calO_u=\calO_{\bar d} \cong 
\frac{\SU(3)_{\rm C} \x \U(1)_{\rm Y}/\bbZ_3}
{\SU(2)_{\rm C}\x \U(1)_{\rm Y-2C}/\bbZ_2}.
\eeq

Approximating the $u$ and $\bar d$ monopoles by ones with the same magnetic
charges in the symmetry breaking
$\SU(4)\rightarrow \SU(3)_{\rm C}\x \U(1)_{\rm Y}/\bbZ_3$ leads to the
gauge orbits 
\beq
\calO_2 = \calO_4 \cong \frac{\SU(3)_{\rm C} \x \U(1)_{\rm Y}/\bbZ_3}
{\SU(2)_{\rm C}\x \U(1)_{\rm Y-2C}/\bbZ_2}.
\eeq
Therefore the gauge orbits of the $u$ and $\bar d$ monopoles and multiplets
are the same. 

\subsubsection{Gauge Freedom of the $(\bar\nu, \bar e)$ Monopole}

The $(\bar\nu, \bar e)$ multiplet transform as an isospin ${\bf 2}$ doublet
with some hypercharge. Therefore its gauge orbit is
\beq
\calO_{(\bar\nu,\bar e)} \cong 
\frac{\SU(2)_{\rm I} \x \U(1)_{\rm Y}/\bbZ_2}
{\U(1)_{\rm Q}},
\eeq
where $\U(1)_{\rm Q}$ lies diagonally between the isospin and hypercharge
groups. Note an interesting equivalence between this gauge orbit and the
electroweak vacuum manifold; this occurs because the associated scalar 
doublet has the same representation as $(\bar\nu, \bar e)$.

Approximating the $(\bar\nu, \bar e)$ monopole by one with the same
magnetic charges in 
$\SU(3)\rightarrow \SU(2)_{\rm I}\x \U(1)_{\rm Y}/\bbZ_2$ then leads
to a gauge orbit  
\beq
\calO_3 \cong \frac{\SU(2)_{\rm I} \x \U(1)_{\rm Y}/\bbZ_2}
{\U(1)_{\rm Q}},
\eeq
which is the same as the $(\bar\nu,\bar e)$ multiplet. 

\subsubsection{Gauge Freedom of the $\bar e$ Monopole}

As the $\bar e$ particle is only charged under hypercharge it has the rather
trivial gauge orbit 
\beq
\calO_{\bar e} \cong \U(1)_{\rm Y}.
\eeq
Similarly, the $\bar e$ monopole is approximated by one with the same
magnetic charge in $\SU(2)\rightarrow \U(1)_{\rm Y}$; giving a gauge
orbit 
\beq
\calO_6 \cong \U(1)_{\rm Y}.
\eeq
This is the same as the $\bar e$ particle.

\section{Spherical Symmetry}
\label{sec:sph}

It is important to understand that all of the monopoles in Georgi-Glashow
SU(5) gauge unification have no intrinsic angular momentum. 

When constructing a dual standard model from SU(5) monopoles 
this appears somewhat problematic because all elementary particles have spin. 
However, what this is really
saying is that SU(5) gauge unification on its own is insufficient to produce a
consistent dual standard model. Because of this it is desirable to understand
the angular momentum properties of SU(5) monopoles; with these properties 
hopefully indicating where to proceed. 

When does a monopole have no intrinsic angular momentum? The accepted
understanding~\cite{Wilk77} is that they are spherically symmetric under
\beq
\vec R = \vec L + \vec T,
\eeq
where $\vec L= -\vec r\wedge\vec\nabla$ generates spatial rotations and 
$\vec T$ generates some $\SU(2)_T$ subgroup of $\SU(5)$. This is saying that
a spatial rotation of a spherically symmetric monopole can always be undone
through an internal gauge rotation of $\SU(2)_T$. That is, every spatial
rotation $S$ is equivalent to an internal rotation $\Om(S)\in \SU(2)_T$
\bea
\label{sph1}
\Phi(\bm r)&=&\Ad[\Om(S)]\, \Phi(S^{-1}\bm r),\\
\label{sph2}
B_i(\bm r)&=&\Ad[\Om(S)]\,S_{ij} B_j(S^{-1}\bm r).
\eea
Here $S(\al,\be,\ga)$ can be defined through its Euler angles, in
which case $\Om(S)$ is $e^{-i\al T_3/2} e^{-i\be T_2/2} e^{-i\ga T_3/2}$.

The spherical symmetry of each SU(5) monopole in
table~\ref{tab1} is now examined using methods first developed by Wilkinson 
and Goldhaber~\cite{Wilk77}. Some of this treatment relates to 
ref.~\cite{miya}.

\subsection{Spherically Symmetric $(u,d)$ Monopoles}
\label{sec-spin0}

A simple illustration of spherical symmetry is provided by the 
$(u,d)$ monopole. The task is to show (\ref{sph1},\ref{sph2}) holds.

Fortunately this monopole has already been expressed in a 
spherically symmetric gauge: the radial gauge (\ref{hoof})  
\beq
\label{h}
\Phi(\bm r) \sim \Ad[\Om(\rhat)]\Phi_0,\hspace{1em}
\bm B (\bm r) \sim \frac{1}{2g}\frac{\rhat}{r^2}\Ad[\Om(\rhat)]M,
\eeq
with $\Om(\rhat)=e^{-i\varphi T_3/2}e^{i\vartheta T_2/2}e^{-i\varphi T_3/2}$
and $\vec T$ defined in (\ref{emb}). 
Then the action of $e^{iT_3\chi}$ upon (\ref{h}) is equivalent
to 
\beq
\Om(\rhat) \mapsto e^{-iT_3\chi}\,\Om(\rhat)\,e^{iT_3\chi}.
\eeq
The point being that this takes $\varphi\mapsto\varphi+\chi$, which is a
spatial rotation around the $z$-axis. The demonstration of spherical symmetry
is then completed by a similar calculation about any other axis. 

\subsection{Spherical Symmetry of the Other Monopoles}
\label{sec:spinother}

Unlike the $(u,d)$ monopole the other SU(5) monopoles in table~\ref{tab1} are
generally fairly complicated. Fortunately there are some simple
criteria for describing their spherical symmetry.

Wilkinson and Goldhaber have constructed a general set of spherically
symmetric monopoles that satisfy (\ref{sph1},\ref{sph2}) for magnetic
generators that decompose into~\cite{Wilk77} 
\beq
\label{sphcon1}
\half M = I_3 - T_3.
\eeq
Here $I_3$ and $T_3$ are elements of two su(2) algebras, whose  generators
$\vec I$ and $\vec T$ are constrained under
\beq
\label{sphcon2}
[\vec I, \Phi_0] = 0,\hspace{3em} [\vec I, M]= 0.
\eeq
These criteria describe the spherical symmetry of all SU(5) monopoles, in a
necessary and sufficient way~\cite{wein84}.

Before discussing these monopoles it will be useful to quickly interpret
the meaning of these conditions:

\noindent (i)
That $\vec I$ commutes with $\Phi_0$ specifies the embedding of the
associated SU(2) group to be contained within the residual symmetry. 

\noindent (ii)
The second constraint is a little more subtle. Later it will be revealed that
this allows the generators $\vec I$ to be globally defined around the monopole.

To see how this spherical symmetry emerges it will be necessary to construct
the asymptotic form of the monopoles satisfying (\ref{sphcon1}) and
(\ref{sphcon2}). The key point is that there is then a gauge transformation
that takes the unitary gauge configuration (\ref{B_top}) to a non-singular
radial form. In that gauge 
\beq
\label{sphla}
\Phi(\bm r) \sim \Ad[\La(\rhat)]\Phi_0,\hspace{1em}
\bm B (\bm r) \sim 
\frac{1}{2g}\frac{\rhat}{r^2} \Ad[\La(\rhat)]M,
\eeq
with $\La(\rhat)=\Om(\rhat)\om^{-1}(\rhat)$, where $\Om(\rhat)$ is as
(\ref{h}) and $\om(\rhat)$ is similarly defined in $\vec I$ as 
$e^{-i\varphi I_3/2}e^{i\vartheta I_2/2}e^{-i\varphi I_3/2}$.

Although, by (\ref{sphcon2}), $\om(\rhat)$ acts trivially upon both $\Phi_0$
and $M$ it is useful for constructing features relating
to the angular momentum. For the time being note that 
$\La$ could be replaced by $\Om$ in (\ref{sphla}) if desired. 

That (\ref{sphla}) is spherically symmetric can be seen through the relation 
$\Ad[\Om(S)\Om(\rhat)]X=\Ad[\Om(S\rhat)]X$, which holds providing $X$ commutes
with $M$~\cite{Wilk77}. Thus $\Om(S)$ acts on the asymptotic fields 
(\ref{sphla}) as
\bea
\Phi(\bm r) &\mapsto& \Ad[\Om(S)] \Phi(\bm r) = \Ad[\La(S\rhat)]\Phi_0,\\
\bm B(\bm r) &\mapsto& \Ad[\Om(S)]\bm B(\bm r)
=\frac{1}{2g}\frac{\rhat}{r^2} \Ad[\La(S\rhat)]M;
\eea
satisfying the definition~(\ref{sph1},\ref{sph2}) of spherical symmetry.

It is now a fairly simple task to construct the relevant generators of the
spherically symmetric SU(5) monopoles. Using table~\ref{tab1} and some trial
and error convinces one that the generators $I_3$ and $T_3$ for a monopole
with magnetic generator $M$ are:  

\begin{table}[h]
\caption{Generators of spherically symmetric monopoles.}
\begin{eqnarray*}
\begin{array}{|c|c|c|c|}
\hline
\vspace*{-.35cm}  &&& \\
 & {\rm diag}\ \half M & {\rm diag}\ I_3 & {\rm diag}\ T_3 \\
\hline
\vspace*{-.35cm}  &&& \\
(u,d) & (0,0,\half,$-$\half,0) & - & (0,0,$-$\half,\half,0) \\
\vspace*{-.35cm} &&& \\
\bar{d} & (0,\half,\half,$-$\half,$-$\half) & - & 
(0,$-$\half,$-$\half,\half,\half) \\
\vspace*{-.35cm}  &&& \\
(\bar{\nu}, \bar{e}) & (\half,\half,\half,$-$1,$-$\half) & 
(0,$-$\half,\half,0,0) & ($-$\half,$-$1,0,1,\half)  \\
\vspace*{-.35cm} &&& \\
u & (\half,\half,1,$-$1,$-$1) & 
(\half,$-$\half,0,0,0) & (0,$-$1,$-$1,1,1) \\
\vspace*{-.35cm} &&& \\
\bar{e} & (1,1,1,$-$\ts{\frac{3}{2}},$-$\ts{\frac{3}{2}}) & 
($-$1,0,1,$-$\half,\half_{\ }) & ($-$2,$-$1,0,1,2) \\
\hline
\end{array}
\end{eqnarray*}
\vspace*{-0.5cm}
\label{tab2}
\end{table}

However, there appears to be a problem with the $u$ monopole: it's $T_3$
eigenvalues $(0,$-$1,$-$1,1,1)$ do not correspond to an SU(2) representation,
which violates the condition below (\ref{sphcon1}). Because of this 
the $u$ monopole is not spherically symmetric.

What does this imply about the $u$ monopole's structure? Note that the
solution almost appears to be spherically symmetric; for instance if the SU(5)
group is enlarged to SU(6) then $T_3$ becomes an SU(2) generator. 
The most likely implication is a small magnetic dipole moment on the $u$
monopole. Certainly the loss of spherical symmetry is qualitatively different
to the angular momentum discussed in the latter parts of this article. However
further study is required to fully elucidate the $u$ monopole's form. 

\subsection{Spherically Symmetric Monopoles with Non-Topological Charge}

In addition to the above monopoles one can also consider their counterparts
with additional non-topological charge. These are generally expected to
be of higher energy, because they have larger magnetic charges.

Some examples of spherically symmetric monopoles having extra non-topological
colour charge are:

\begin{table}[h]
\caption{Non-Topological Monopole Charges.}
\begin{eqnarray*}
\begin{array}{|c|c|c|c|}
\hline
\vspace*{-.35cm}  &&& \\
 & {\rm diag}\ \half M & {\rm diag}\ I_3 & {\rm diag}\ T_3 \\
\hline
\vspace*{-.35cm}  &&& \\
u^* & (1,1,0,$-$1,$-$1) & ($-$\half,\half,0,$-$\half,\half) & 
(\ts{\frac{3}{2}},\half,0,$-$\half,$-$\ts{\frac{3}{2}}) \\
\vspace*{-.35cm} &&& \\
u^{**} & (0,0,2,$-$1,$-$1) & (0,$-$\half,\half,0,$-$\half,\half) & 
(0,\half,\ts{\frac{3}{2}},$-$\half,$-$\ts{\frac{3}{2}}) \\
\vspace*{-.35cm}  &&& \\
\bar{e}^* & (\half,\half,2,$-$\ts{\frac{3}{2}},$-$\ts{\frac{3}{2}}) & 
(\half,$-$\half,0,\half,$-$\half) & (1,0,2,$-$1,$-$2) \\
\vspace*{-.35cm} &&& \\
\bar{e}^{**} & 
(\ts{\frac{3}{2}},\ts{\frac{3}{2}}$-$,0,
$-$\ts{\frac{3}{2}},$-$\ts{\frac{3}{2}}) & ($-$\half,\half,0,$-$\half,\half)
& (2,1,0,$-$1,$-$2) \\
\hline
\end{array}
\end{eqnarray*}
\vspace*{-0.5cm}
\label{tab3}
\end{table}

\noindent
There are in fact an infinite tower of non-topological magnetic charges on each
monopole. Many of these will be spherically symmetric.

\section{SU(5) Dyons}

The rest of this article is mainly concerned with SU(5) dyons, which 
have electric charge in addition to their original magnetic charge. This
charge is crucial to their nature and dynamics; for instance many of these 
dyons have an intrinsic angular momentum. In some cases this 
angular momentum is half integer. 

That these dyons may have angular momenta of one-half is really
very encouraging for the construction of a dual standard model from
SU(5) solitons. As the SU(5) monopoles stand they cannot be dual to the 
elementary particles because they have no intrinsic angular momenta; but 
by combining them with electric charges the required solitons can be 
constructed. Indeed it has been shown that each SU(5) monopole has a dyonic
counterpart with one-half angular momentum~\cite{vachspin}; these also have
interesting duality properties~\cite{steer}. 

Thus, for constructing a dual standard model, an important consideration 
is the spectrum and properties of SU(5) dyons. This motivates a
detailed discussion of these dyons, which naturally splits into three parts:

\noindent 
(i) {\em Scalar boson-monopole composites}: this is the subject of
sec.~(\ref{sec:spin}) and is conceptually the simplest case. The inclusion
of extra scalar fields allows their quanta to bind to the monopoles, 
giving dyons. Many of these have intrinsic angular momentum. It is these
dyons that have been discussed in refs.~\cite{vachspin,steer}.

\noindent 
(ii) {\em Monopole gauge excitations}: these are the subject of
sec.~(\ref{sec:gspin}) and arise from the monopole's gauge
freedom. Essentially an electric field is produced by internal motion of the
monopole through its gauge equivalent states. Again these dyons appear to 
have angular momentum.

\noindent 
(iii) {\em Effects of a $\theta$ vacuum}: this is the subject of
sec.~(\ref{sec:theta}). A theta vacuum changes the definition of the 
electric field from the Noether charges, which effectively induces 
electric charge on the monopole. Whilst this induced charge does not induce
angular momentum it does have a central effect on the nature of the dyon
spectrum.

Much of the discussion in these three sections relies on similar background
material. This is discussed over the next few subsection and hopefully
provides a useful introduction to the properties of dyons.

\subsection{Dyon Configurations}

The first topic of concern is the definition of the electric and magnetic
fields around a dyon. These may be considered in a unitary gauge 
\beq
\label{dyonsu}
\Phi \sim \Phi_0,\hspace{1em}
\bm E \sim g \frac{\rhat}{4\pi r^2}Q,\hspace{1em}
\bm B \sim \frac{1}{2g}\frac{\rhat}{r^2} M.
\eeq 
Particular attention is paid to the magnetic and electric charge 
normalisations, which will be important later. 

Before starting note the
unitary gauge configuration (\ref{dyonsu}) has a Dirac string, which must
be unobservable by the electric charge to be well defined. This constrains 
$Q$ and $M$ through a non-Abelian Dirac condition
\beq
\label{dcon}
\tr\,QM \in \bbZ,
\eeq
which follows from projecting the non-Abelian theory into the Abelian
$\U(1)_M$ subtheory.

\subsubsection{Magnetic Charge}

As discussed in sec.~(\ref{sec:dsm}), the magnetic generator is defined
through the condition $\exp(i2\pi M)=1$. This specifies
the individual magnetic charges
\beq
M = m_{\rm C}T_{\rm C}+m_{\rm C'}T_{\rm C'}
+m_{\rm I}T_{\rm I}+m_{\rm Y}T_{\rm Y},
\eeq
with respect to properly normalised colour, isospin and hypercharge
generators (\ref{gen}). The relevance of these magnetic charges 
can be seen by extracting the individual magnetic fields
from the full magnetic field 
$\bm B=\bm B_\rmC T_\rmC +\bm B_{\rmC'}T_{\rmC'}
+\bm B_\rmI T_\rmI+\bm B_\rmY T_\rmY$.

Of particular interest are the monopole's colour and isospin magnetic 
charges. In the normalisation of (\ref{gen}) these take the non-zero values
\bea
(m_\rmC, m_{\rmC'})=\{(-\surd\ts\frac{1}{3},\pm\surd\ts\frac{1}{2}),
(\surd\ts\frac{2}{3},0)\},\hspace{1em}
m_{\rm I}=\pm\surd\ts\frac{1}{2}.\nn
\eea
Note that these values are identical to the eigenvalues of the colour and
isospin generators, for example in table~\ref{tab4} below. This is the
principal reason for taking this normalisation. 

\subsubsection{Electric Charge}

To treat electric and magnetic charge on a similar footing the
dyon's electric generator is defined as
\beq
Q=
q_{\rm C}T_{\rm C}+q_{\rm C'}T_{\rm C'}+q_{\rm I}T_{\rm I}+q_{\rm Y}T_{\rm Y},
\eeq
with each $q$ an individual electric charge. In the above normalisation to
tr$T^2=1$ the electric charges are related to the Noether current in the usual
way; with each charge an eigenvalue of the relevant generator.

For example a fundamental $\bm 5$ scalar field $H$ has electric charges that 
are eigenvalues of $T_{\rm C}$, $T_{\rm C'}$, $T_{\rm I}$ and $T_{\rm Y}$:

\begin{table}[ht]
\caption{Scalar Charges.}
\begin{eqnarray*}
\begin{array}{|c|ccc|c|}
\hline
&(q_{\rm C},q'_{\rm C})&q_{\rm I}&q_{\rm Y}& {\rm diag}\ Q_i \\
\hline
\vspace*{-.35cm}  &&&& \\
H_1 & ($-$\surd\ts\frac{1}{3},\surd\half) & 0 & \ \ \surd\!\frac{2}{15}&
(\fourfifth,$-$\fifth,$-$\fifth,$-$\fifth,$-$\fifth)\\
\vspace*{-.35cm}  &&&& \\
H_2 & ($-$\surd\ts\frac{1}{3},$-$\surd\half) & 0 &\ \ \surd\!\frac{2}{15}&
($-$\fifth,\fourfifth,$-$\fifth,$-$\fifth,$-$\fifth) \\
\vspace*{-.35cm}  &&&& \\
H_3 & (\surd\ts\frac{2}{3},0) & 0 & \ \ \surd\!\frac{2}{15}&
($-$\fifth,$-$\fifth,\fourfifth,$-$\fifth,$-$\fifth) \\
\vspace*{-.35cm}  &&&& \\
H_4& 0 & \surd\half & \ \ $-$\frac{3}{2}\surd\!\frac{2}{15}&
($-$\fifth,$-$\fifth,$-$\fifth,\fourfifth,$-$\fifth) \\
\vspace*{-.35cm}  &&&& \\
H_5 & 0 & $-$\surd\half & \ \ $-$\frac{3}{2}\surd\!\frac{2}{15}&
($-$\fifth,$-$\fifth,$-$\fifth,$-$\fifth,\fourfifth) \\
\hline 
\end{array}
\end{eqnarray*}
\vspace*{-0.5cm}
\label{tab4}
\end{table}

\noindent
The electric generator of a scalar boson associated with the $H_i$ component
is labelled $Q_i$. Around such a boson is the electric field  
\beq
\bm E = g\frac{\rhat}{4\pi r^2} Q_i.
\eeq
This relates to the classical limit of the Noether current, which for a
single stationary point charge is $j^{\rm cl}_0 = Q \de^3(\bm r)$.

Note that the numerical values of these electric charges are the same 
as the magnetic charges of the monopoles. As stated above, this is due 
to the normalisation in (\ref{gen}). 

\subsection{Global Charge}
\label{secglob}

An important, and initially unexpected, feature of the above dyons 
(\ref{dyonsu}) is that generally their electric $Q$ and magnetic $M$ 
generators are not independent. This feature is central to 
determining the SU(5) dyon spectrum. 

The reason for this dependence between $Q$ and $M$ is a rather elegant
property of gauge theories, where an electric charge may not be globally 
defined around a non-Abelian monopole~\cite{globcol}. It transpires there
is a topological obstruction to defining such electric charge. 

A way to appreciate this global charge is to consider how the 
gauge field is patched around a monopole~\cite{wuyang}. In the unitary 
gauge the asymptotic
magnetic field $\rhat M/2gr^2$ can be defined by two different gauge
potentials, 
\beq
\label{A0N}
{\bm A}_N \sim \frac{1}{2g}
\frac{1-\cos\vartheta}{r\sin\vartheta}M\hat\bm\varphi,
\eeq
with a Dirac string along the negative $z$-axis, or
\beq
{\bm A}_S \sim -\frac{1}{2g}
\frac{1+\cos\vartheta}{r\sin\vartheta}M\hat\bm\varphi,
\eeq
with a Dirac string along the positive $z$-axis. As these potentials define
the same monopole they are gauge equivalent under
$h(\varphi)=\exp(iM\varphi)$. 

The point is that a monopole can be well defined everywhere, with no string,
in the unitary gauge providing the gauge field is patched around the
monopole~\cite{wuyang}. 
Then asymptotic space is split into two patches, say the
north and south hemispheres, with ${\bm A}_N$ and ${\bm A}_S$ well defined on
their respective patches. The constraint that these two patches can be joined
together by a well-defined gauge transformation implies that $h(2\pi)=1$; 
yielding the topological quantisation (\ref{qq}).

What happens if there is an electric charge? Then there is
an additional component to the gauge field 
\beq
A^N_0 \sim \frac{g}{4\pi r}\, Q.
\eeq
But this is only consistent with the patching if on the southern hemisphere
$A_0^S=\Ad[h(\varphi)]A_0^N$, which has a string singularity along the
negative $z$-axis unless 
\beq
\label{globch}
[Q,M]=0.
\eeq
In other words an electric charge is only allowed on a magnetic monopole
if the electric generator is an element of an `allowed' symmetry group.
This group acts trivially on $M$, otherwise the dyon will not have a well
defined gauge field everywhere around it~\cite{globcol,bala}.

For each of the monopoles these allowed symmetry groups are given in
table~\ref{tab5} below.

\begin{table}[h]
\caption{Allowed Gauge Symmetries $H_A$.}
\begin{eqnarray*}
\begin{array}{|c|c|}
\hline
 & H_A \\
\hline
(u,d) & \U(1)_\rmC\x \U(1)_\rmI\x \U(1)_\rmY\x \SU(2)_{\rmC'}/\bbZ_{12}\\
\bar d & \U(1)_\rmC\x\SU(2)_\rmI\x\U(1)_\rmY\x\SU(2)_{\rmC'}/\bbZ_{12}\\
(\bar\nu,\bar e) & \SU(3)_\rmC \x \U(1)_\rmI \x \U(1)_\rmY/\bbZ_6 \\
u & \U(1)_\rmC \x \SU(2)_\rmI \x \U(1)_\rmY \x \SU(2)_{\rmC'}/\bbZ_{12}\\
\bar e & \SU(3)_\rmC \x \SU(2)_\rmI \x \U(1)_\rmY/\bbZ_6 \\
\hline 
\end{array}
\end{eqnarray*}
\vspace*{-0.5cm}
\label{tab5}
\end{table}

\noindent
In each of these there is a quotient by a finite intersection whose
details are not central here.

It is important to realise that
this non-definition of global charge goes deeper. Even the notion
of gauge invariance can fail outside $H_A$, which effectively restricts the 
charge-monopole gauge interactions to within this group~\cite{bala}. 

\subsection{Angular Momentum}
\label{secang}

An important property of these electrically and magnetically charged dyons is
that many appear to have non-trivial intrinsic angular momenta. This
section describes this phenomenon from a classical perspective for dyons 
composed of separate electric and magnetic charges.
Other methods will be used later for evaluating this angular momenta; all
of which are consistent with the results below. 

One way of understanding a dyon's intrinsic angular momentum is through its
non-Abelian electric-magnetic field. Strictly speaking these methods are only
relevant for dyons whose individual components have no angular momenta; indeed
care should be taken when applying them to more complicated situations.
The intrinsic angular momentum is derived from the non-Abelian generalisation 
of the Poynting vector, which gives
\beq
\label{J1}
\bm J = \int {\rm d}^3r\, \tr[ \bm r \wedge (\bm E \wedge \bm B)].
\eeq
Evaluating this for a dyon with magnetic charge centred at the origin and 
electric charge at $\bm r$ yields
\beq
\label{eq:j}
\bm J = \half\, \tr\, QM \,\hat\bm r.
\eeq

A simple proof of (\ref{eq:j}) has been presented by Goddard
and Olive~\cite{mon}. Considering the components of (\ref{J1}) gives
\bea
J_i &=&\int{\rm d}^3r\,\tr[ E_j(\de_{ij}-\hat x_i\hat x_j)\frac{M}{2gr}]\nn\\
&=&
\int{\rm d}^3r\,\tr[ E_j \pderiv{}{x_j}\left(\frac{\hat x_j}{2g}M\right)]\nn\\
&=&-\int{\rm d}^3r\,\tr[\bm\nabla\cdot\bm E\, \frac{\hat x_j}{2g}M],
\eea
which results in (\ref{eq:j}).

\subsection{Angular Momentum and Statistics}
\label{sec:gold}

Application of the Dirac condition (\ref{dcon}) to the
angular momentum (\ref{eq:j}) implies that $J$ must either be an integer or
half-integer. Thus, providing that the usual relation between spin and 
statistics holds, one might expect some dyons to be fermionic even though the
constituents are bosonic. That this is indeed the case
was demonstrated by Goldhaber~\cite{gold:spin}.

To see this consider two dyons $(Q,M)$ at $\pm \bm x$ moving with velocities
$\pm \bm v$. Then the current-gauge interaction is
\beq
\label{cint}
\calH_{\rm int} = -\tr[(gQ\bm v)\cdot(\bm A(\bm x) - \bm A(-\bm x))],
\eeq
with a gauge potential as in (\ref{A0N}). At first sight this is
a very complicated, velocity dependent, interaction. However it can
be simplified by noting
\beq
\tr\,Q(A_\varphi(\vartheta,\varphi) - A_\varphi(\pi-\vartheta,\varphi+\pi)) 
= \frac{1}{ig}(\partial_\varphi\Om)\Om^{-1},
\eeq
where $\Om=\exp(i\varphi\,\tr\, QM)$.

Thus if one considers the two-dyon wavefunction $\Psi(\bm x)$, then the 
complicated interaction (\ref{cint}) can be removed by a gauge transformation
\beq
\Psi(\bm x)\mapsto\Om(\bm x)\Psi(\bm x).
\eeq
Then upon interchange of the dyons $\bm x\mapsto -\bm x$, this gauge 
transformation effects the phase of the wavefunction 
\beq
\Om(-\bm x)=\Om(\varphi+\pi)=\exp(i\pi\tr\,QM)\Om(\bm x).
\eeq
Hence the usual connection between spin and statistics is obtained, where
for integer/half-integer $J$ the wavefunction is symmetric/antisymmetric 
upon dyon interchange.



\section{Dyonic Scalar Boson-Monopole Composites}
\label{sec:spin}

A simple method for constructing dyons from SU(5) monopoles is to 
form composites with electric scalar bosons. Although their angular momenta
can be evaluated from their electric-magnetic fields (\ref{eq:j}), it will be
useful to examine the nature of the resulting solitons from a 
semi-classical viewpoint.

To be specific the scalar bosons are taken to be quanta of a $\bm 5$ scalar 
field $H$. Such a field is directly relevant to SU(5) grand unification, since
it is generally taken to contain the isospin doublet responsible for 
electroweak symmetry breaking. In this context, Lykken and Strominger 
demonstrated the $(u,d)$ monopole to have dyonic counterparts with one-half 
angular momentum~\cite{lyk80}. 

Following their analysis, Vachaspati has examined the angular momenta of the 
scalar boson-monopole composites in the context of a dual standard model.
Using the classical methods of sec.~(\ref{secang}) he verified that each 
SU(5) monopole in table~\ref{tab1} has a dyonic counterpart with one-half 
angular momentum. This gives strong support for using such dyons to 
construct a dual formulation of the standard model.

This section discusses these dyons from the semi-classical
perspective of 't~Hooft and Hasenfratz. Then the properties of the
dyons are described by the quantum mechanics of the scalar bosons in the 
classical SU(5) monopole background. The appropriate Hamiltonian is 
specified by the monopole's gauge potential 
\beq
\hat{\calH}=(1/2m) \bm D^2 + \,V(r),
\hspace{1em}\bm D = i\bm\nabla + \bm A(\bm r)
\eeq
and acts upon the five-component scalar field $H$.

An important feature, which will be discussed again in sec.~(\ref{sec:theta}),
it that a binding potential $V(r)$ is required between the charge and 
monopole. For simplicity this is taken to be spherically symmetric.
It should be noted that such a potential is required to construct the dual 
standard model, since otherwise the dyons are not stable bound states. 

The spherical symmetry of the Hamiltonian $\hat\calH$ implies there
is a conserved angular momentum, of which the $z$-component is
\beq
\label{J}
\hat J_3 = [\bm r \wedge \bm D]_3 + \half\, M.
\eeq
Then the angular momentum of a composite dyon $(Q,M)$ is simply the 
appropriate eigenvalue of $\hat J_3$ for the $H$ eigenstate. For a 
spherically symmetric ground state the angular momentum is then 
determined only by 
\beq
\label{angmom}
\hat{J}_3 H_i(r) = \half M H_i(r) = J_3 H_i(r).
\eeq
By (\ref{qq}) these eigenvalues are either integer or half integer; giving
integer or half-integer angular momenta to the resulting dyons.

It is then a simple task to determine the angular momenta of the different
scalar boson-monopole composites; with the $J_3$ values simply corresponding 
to the relevant eigenvalue of $\half M$. Using table~\ref{tab1} these are:

\begin{table}[h]
\caption{Angular momenta of the $(Q,M)$ scalar 
boson-monopole composites.}
\begin{eqnarray*}
\begin{array}{|c|c|ccccc|}
\hline
& {\rm diag}\ M & Q_1 & Q_2 & Q_3 & Q_4 & Q_5 \\
\hline
(u,d)&(0,0,1,$-$1,0)&0&0&\half&$-$\half&0  \\
\bar{d}&(0,1,1,$-$1,$-$1)&0&\half&\half&$-$\half&$-$\half \\
(\bar\nu,\bar e)&(1,1,1,$-$2,$-$1)&\half&\half&\half&$-$1&$-$\half \\
u&(1,1,2,$-$2,$-$2)&\half&\half&1&$-$1&$-$1 \\
\bar{e}&(2,2,2,$-$3,$-$3)&1&1&1&$-$\frac{3}{2}&$-$\frac{3}{2} \\
\hline
\end{array}
\end{eqnarray*}
\vspace*{-0.5cm}
\label{tab6}
\end{table}

\noindent
These are consistent with the angular momentum~(\ref{eq:j}). Note the
$u$ monopole has some subtlety, as described in the last paragraph
of sec.~(\ref{sec:spinother}).

Table~\ref{tab6} does not constitute the complete spectrum of dyons. In
addition there can also be quanta of $H$ bound to anti-monopoles, or 
anti-particles of $\bar H$ bound to monopoles/anti-monopoles. To simplify their
classification note that this reflects an underlying parity and
charge-parity symmetry,  
\bea
\label{p}
\calP:\, (Q, M) \mapsto (Q, -M),\hspace{3em}
\calP:\, J_3 \mapsto -J_3;\\
\label{cp}
\calC\calP:\, (Q, M) \mapsto (-Q, M),\hspace{3em}
\calC\calP:\, J_3 \mapsto -J_3.
\eea
Parity takes a monopole to its anti-monopole, whilst charge-parity
takes an electric charge to its anti-charge; both reverse spin. This can be
conveniently represented in the following figure, which is based on a
discussion in ref.~\cite{vach95}.

\begin{figure}[h]
\begin{center}
\epsfxsize=16em \epsfysize=10em \epsffile{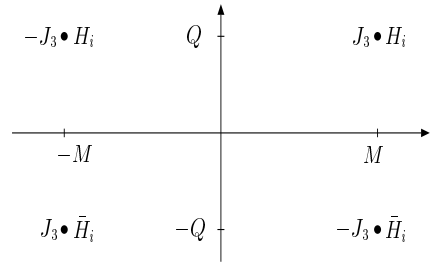}
\end{center}
\caption{$H_i$ scalar bosons bound to either a monopole or 
an anti-monopole.}
\label{fig1}
\end{figure}

\noindent
Here $\calP$ is a reflection about the $Q$ axis, whilst $\calC\calP$
reflects about the $M$ axis.

However the above methods do not appear to produce an $\bar e$ dyon with 
one-half angular momentum. If such a state does not exist this would seriously
jeopardise the construction of a dual standard model from SU(5) solitons.
Fortunately, there are a couple of solutions:

\noindent (i)
It appears that two quanta of $H$ on an $\bar e$ monopole could have one-half 
angular momentum~\cite{vachspin}. 

\noindent (ii)
An $\bar e^*$ monopole with extra non-topological colour charge is spherically
symmetric and has scalar excitations with angular momentum one-half.

\begin{table}[h]
\caption{Angular momenta of the $\bar e^*$ dyons.}
\begin{eqnarray*}
\begin{array}{|c|c|ccccc|}
\hline
& {\rm diag}\ M & Q_1 & Q_2 & Q_3 & Q_4 & Q_5 \\
\hline
\vspace*{-0.35cm} && \\
\bar e_L^* & (1,1,4,$-$3,$-$3) &
\half & \half & 2 & $-$\frac{3}{2} & $-$\frac{3}{2} \\
\hline 
\end{array}
\end{eqnarray*}
\vspace*{-0.5cm}
\label{tab7}
\end{table}

\section{Dyons from Monopole Gauge Excitations}
\label{sec:gspin}

This section discusses dyons whose electric charge originates within the 
properties of 
the semi-classical electric field around a monopole. These have a similar 
nature to the dyons discussed in sec.~(\ref{sec:spin}); 
in particular it appears that they may also possess intrinsic angular 
momentum.
Therefore when constructing a dual standard model the dyons formed by 
combining monopoles with scalar bosons or gauge excitations
are equally relevant. 

The plan of this section is to firstly illustrate how a classical electric
field can arise around a monopole. This is then quantised through
semi-classical methods. Finally the angular momentum of such dyons 
is discussed.

\subsection{Classical Dyon Charge}
\label{sec:class}

To start it will prove useful to examine the nature of the magnetic and
electric fields around a dyon within a classical field theory context.
The following analysis is essentially a
non-Abelian generalisation of the Julia-Zee dyon~\cite{julia75}.

The task is to find a long range component $A_0$ that solves the field
equations in the background of the monopole. Taking a unitary gauge and
assuming the electric field to be purely radial allows the field equations to
be written, asymptotically, 
\beq
\label{didi}
D_i D_i A_0 - [\Phi_0,[\Phi_0,A_0]] =0,
\eeq
where $D_i$ is the covariant derivative in the background of a monopole and,
to be specific, the spatial components of the gauge potential are as in
(\ref{A0N}). This implies that the leading order contribution to $A_0$
satisfies   
\beq
[A_0,\Phi_0]=[A_0,M]=0,
\eeq
for which (\ref{didi}) becomes a Laplace equation. Therefore in the unitary
gauge the classical dyon configuration is
\bea
\label{dyfield}
\Phi(\bm r)\sim\Phi_0,\hspace{2em}
\bm A(\rhat) \sim \frac{1}{2g}
\frac{1-\cos\vartheta}{r\sin\vartheta}M\hat\bm\varphi,
\nn\\ A_0(\bm r) \sim \frac{g}{4\pi r}Q,\hspace{4em}
\eea
with $[Q, \Phi_0]=[Q, M]=0$, in agreement with the constraint~(\ref{globch})
on global charge. 

Because this analysis was classical the magnitude of the electric charge 
can take a
continuum of values. Clearly this is not consistent quantum mechanically; for
instance it violates Dirac's condition and also the angular momentum
(\ref{eq:j}) is not constrained to be 
integer or half-integer. Consistent values are 
obtained only upon proper quantum mechanical treatment. 

This charge can also be interpreted in a slightly different way. Upon 
performing a time dependent gauge transformation
\bea
\Phi(\bm r,t)&=&\Ad[U(r,t)] \Phi(\bm r),\\
\bm A(\bm r,t)&=&\Ad[U(r,t)] \bm A(\bm r),
\eea
the profile (\ref{dyfield}) can be expressed in an $A_0=0$ gauge providing
\beq
\label{udot}
\dot U U^{-1} = \frac{g}{4\pi r}Q.
\eeq
Hence the dyon can be thought of as a monopole rotating in internal space
under the action of $U(r,t)$. This motion is quantised to discretise charge.

However when quantising such internal motion only the non-trivial actions of
$U$ are relevant. For each SU(5) monopole the subgroup of $H_A$ that acts
non-trivially is:  

\begin{table}[h]
\caption{Allowed and Effective Gauge Symmetries.}
\begin{eqnarray*}
\begin{array}{|c|c|}
\hline
\vspace*{-0.35cm} & \\
& H_A^{\rm eff} \\
\hline
(u,d) & \U(1)_M \\
\bar d & \U(1)_{\rmC} \x \SU(2)_\rmI \x \U(1)_\rmY/\bbZ_6 \\
(\bar\nu,\bar e) & \SU(3)_\rmC \x \U(1)_\rmI \x \U(1)_\rmY/\bbZ_6 \\
u & \U(1)_\rmC \x \SU(2)_\rmI \x \U(1)_\rmY \x \SU(2)_{\rmC'}/\bbZ_{12}\\
\bar e & \SU(3)_\rmC \x \SU(2)_\rmI \x \U(1)_\rmY/\bbZ_6 \\
\hline 
\end{array}
\end{eqnarray*}
\vspace*{-0.5cm}
\label{tab8}
\end{table}

\subsection{Quantisation of the Gauge Excitations}
\label{secq}

A semi-classical quantisation of the charge values (\ref{dyfield}) can
be achieved through the method of collective coordinates. Then the 
quantum state is described by a wavefunction over the monopole's internal 
degrees of freedom, with the quantised excitations
described by the charge eigenstates.

Much of this section follows the work of Dixon~\cite{dixon84}, although
Guadagnini's quantisation of skyrmions~\cite{guad84} is also relevant.   

The quantisation of the monopole's internal degrees of freedom can be
considered over the symmetry group $H_A^{\rm eff}$, which generates the
motion. Then the quantum excitations are described by a wavefunction
$\psi(U)$, written as a sum over representations
\beq
\label{psi}
\psi(U) = \sum_r \psi^r(U) = 
\sum_r \sum_{ij} c_{ij}^r D_{ij}^r(U),\ \ U\in H_A^{\rm eff},
\eeq
where $D_{ij}^r$ are the matrix elements in the $r$-representation of 
$H_A^{\rm eff}$. Each of the terms in (\ref{psi}) corresponds to a particular
gauge excitation. The charge of these excitations is then determined by the
action of the gauge group, namely
\beq
Q\mapsto \Ad[h]\,Q\ \ \Rightarrow\ \ \ U\mapsto h\,U.
\eeq
The charges therefore fall into representations of the $H_A^{\rm eff}$.

The specific charges of the gauge excitations are constructed from the
relevant charge operators. These are analogues of the momentum operator 
$\hat p = -i\partial_x$, acting instead on the internal motion 
\beq
\hat\calO\, \psi(U) = \psi(U + \de U) - \psi(U) = \psi(\de U),
\eeq
where $\de U$ is an infinitesimal transformation of the symmetry associated
with the conserved charge. This gives the charge operators
\beq
\label{QT}
\hat Q_T\, \psi(U) = \ \psi(T U)\  = q_T \psi(U).
\eeq
Its eigenfunctions are specified by $c_{ij}^r$ and $U$.  

Then the individual charge eigenstates correspond to the representations of
$H_A^{\rm eff}$ in table~\ref{tab8}, which are
\beq
D^r(U) = D(U_\rmC) D(U_\rmI)\, U_\rmY^r,\ \ \ \ 
U_\rmY^r=e^{iq_\rmY T_\rmY \theta},
\eeq
with $r$ an integer. The first few terms of $\psi(U)$ in (\ref{psi})
determines the charge associated with each excitation, which, for suitable
eigenfunctions, are the eigenvalues  
\beq
d^r(T_\rmC) U = q_\rmC\, U,\ \ \ 
d^r(T_\rmI) U = q_\rmI\, U.
\eeq
This gives the spectrum:

\begin{table}[h]
\caption{Dyonic Gauge Excitations.}
\begin{eqnarray*}
\begin{array}{|c|ccc|c|c|}
\hline
\vspace*{-0.35cm} &&&&& \\
D(h) & q_\rmC & q_\rmI & q_\rmY & {\rm diag}\ Q & {\rm allowed\ on}\\
\hline
\vspace*{-0.35cm} &&&&& \\
h_\rmC h_\rmI h_\rmY 
& \surd\frac{2}{3} &\surd\frac{1}{2} &\frac{1}{3}\surd\!\frac{2}{15}
& (0,0,1,$-$1,0) &{\rm all} \\
\vspace*{-0.35cm} &&&&& \\
h_\rmC^\dagger h_\rmY^2
& $-$\surd\frac{2}{3} & 0 & \frac{2}{3}\surd\!\frac{2}{15} & 
(0,1,1,$-$1,$-$1) & \bar d, (\bar\nu, \bar e), u, \bar e\\ 
\vspace*{-0.35cm} &&&&& \\
h_\rmI h_\rmY^3
& 0 & \surd\frac{1}{2} & 1\surd\!\frac{2}{15}  & 
(1,1,1,$-$2,$-$1) & (\bar\nu, \bar e), u, 
\bar e\\
\vspace*{-0.35cm} &&&&& \\
h_\rmC h_\rmY^4
& \surd\frac{2}{3} & 0 &\frac{4}{3}\surd\!\frac{2}{15}&
(1,1,2,$-$2,$-$2) & (\bar\nu, \bar e), u, \bar e \\ 
\vspace*{-0.35cm} &&&&& \\
h_\rmC^\dagger h_\rmI h_\rmY^5
& $-$\surd\frac{2}{3} &\surd\frac{1}{2} &\frac{5}{3}\surd\!\frac{2}{15} &
(1,2,2,$-$3,$-$2) & (\bar\nu, \bar e), u, \bar e \\
\vspace*{-0.35cm} &&&&& \\
h_\rmY^6
& 0 & 0 & 2\surd\!\frac{2}{15} & 
(2,2,2,$-$3,$-$3) & (\bar\nu,\bar e), u, \bar e \\
\hline 
\end{array}
\end{eqnarray*}
\vspace*{-0.5cm}
\label{tab9}
\end{table}

\noindent
There are charges other than those in table~\ref{tab9}. However these are
not central to the following discussion; being highly charged and therefore 
more energetic.   

It is interesting that all electric generators satisfy
$\exp(i\,2\pi Q)=1$, and therefore take the same values as the magnetic 
generators of the monopoles. This is because both the monopoles and gauge
excitations occur in specific representations of $H_\sm$.  

\subsection{Spherically Symmetric Gauge Excitations}
\label{secsp}

In sec.~(\ref{sec:spinother}) the monopoles with no intrinsic angular
momentum were expressed in spherically symmetric way 
\beq
\label{eqla}
\Phi(\bm r) \sim \Ad[\La(\rhat)]\Phi_0,\hspace{1em}
\bm B(\bm r) \sim \frac{1}{2g}\frac{\rhat}{r^2}\Ad[\La(\rhat)]M.
\eeq
Then a gauge transformation by $\Om(S)$ simply takes
\beq
\Phi(\bm r) \mapsto \Phi(S\bm r),\hspace{1em}
B_i(\bm r) \mapsto S_{ij}^{-1}B_j(S\bm r),
\eeq
so that a spatial rotation is equivalent to a gauge transformation.

So what happens when there is an electric charge on the
monopole? In the gauge of (\ref{eqla}) the electric field is 
\beq
\bm E(\bm r) \sim g\frac{\rhat}{4\pi r^2}\Ad[\La(\rhat)]Q,
\eeq
which should be non-singular for those charges allowed on the
monopole. From this a gauge transformation by $\Om(S)$ takes 
\beq
\label{omQ}
\bm E(\bm r) \mapsto g\frac{\rhat}{4\pi r^2} 
\Ad[\La(S\rhat)]\,\Ad[\om(S)]\, Q.
\eeq
Therefore a gauge excitation is spherically symmetric with no angular 
momentum if
\beq
\label{sdycon}
[\vec I, Q]=0.
\eeq

The determination of the spherically symmetric gauge excitations is generally
treated on a case by case basis. However there are some gauge
excitations, for instance when the electric charge is proportional to the 
magnetic charge $(\pm M, M)$, when condition (\ref{sdycon}) is always 
satisfied. This has relevance to the induced charge from a theta vacuum 
discussed in sec.~(\ref{sec:theta}).

\subsection{Monopole Gauge Excitations with Internal Angular Momenta}

On examining the spectrum of monopole gauge excitations in table~\ref{tab9} 
it transpires that not all are spherically symmetric,
since many violate condition (\ref{sdycon}). For illustration some 
particular examples are given table~\ref{tab10} below. That these dyons are
not spherically symmetric suggests they have internal angular momenta, 
just like the dyons in sec.~(\ref{sec:spin}).

Unlike the composite scalar boson-monopole dyons 
the determination of the intrinsic angular momentum of the monopole gauge
excitations is complicated by the internal structure of the gauge excitation.
This may be seen by naively applying (\ref{eq:j}) to the spherically symmetric
gauge excitations of sec.~(\ref{secsp}), where non-sensical results are 
generally obtained.

However it does seem that some of these monopole gauge
excitations are fermionic. This is because Goldhaber's arguments of
sec.~(\ref{sec:gold}) appear to apply; then many dyons, for instance
those in table~\ref{tab10}, do seem to have anti-symmetric statistics. 
Unless the spin-statistics relation is violated then these
should have half-integer angular momenta.

Unfortunately the angular momenta of these gauge excitations is an area
that has not been fully examined. At present there is only one treatment,
by Dixon~\cite{dixon84}, that has studied their properties. He used the 
semi-classical methods of sec.~(\ref{secq}) to define their angular
momenta and found an agreement with the spin-statistics relation.

In this section a classical approach will be used 
to model the gauge excitations. For the dyons in table~\ref{tab10} below
this gives compatible results with Dixon's approach.

To model these gauge excitations consider a composite of a spherically 
symmetric gauge excitation and a gauge boson. Then the
electric charge generator can be split into two components 
\beq
\label{mod}
Q = Q_0 + Q_s,\hspace{2em} [\vec I, Q_0]=0,
\eeq
where $Q_0$ contributes no angular momentum.  Taking $Q_s$ to be
${\rm diag}(1,$-$1)$ embedded along the diagonal gives a single
gauge boson charge eigenstate. 

The configurations in (\ref{mod}) can have the same electric and magnetic
charges $(Q,M)$ as the gauge excitations. Therefore they should have the 
same statistics. Their angular momenta can also be checked to be
compatible with Dixon's results for those in table~\ref{tab10} 
below. This suggests that they model the gauge 
excitations. Whether this is true in general, or just for specific
cases, such monopole-gauge boson composites should be present in the 
SU(5) monopole theory.

The angular momentum of these composites is then
determined from (\ref{eq:j}) to be
\beq
J_3=\half \tr\,{Q_sM} + s_3,
\eeq 
which appears to be degenerate in the spin value $s_3$ of the gauge boson. 
This degeneracy is lifted by the spin-magnetic field interaction
\beq
\calH_{\rm int}=\tr\,Q\bm s\cdot\bm B.
\eeq
Then the dyon states with least energy have angular momenta
\beq
J_3=\left\{ \begin{array}{c}
\half \tr\,{Q_sM} -1,\ \ \ \tr\,Q_sM\geq 0,\\
\half \tr\,{Q_sM} +1,\ \ \ \tr\,Q_sM\leq 0,\end{array}\right.
\eeq

This construction is illustrated 
with the following electric colour excitations:

\begin{table}[h]
\caption{Dyonic Gauge Excitations with one-half angular momentum.}
\begin{eqnarray*}
\begin{array}{|c|c|c|c|}
\hline
{\rm dyon}& {\rm diag}\ M & {\rm diag}\ Q & J_3 \\
\hline 
\vspace*{-0.35cm} &&& \\
(\bar\nu,\bar e) & (1,1,1,$-$2,$-$1) & (0,1,1,$-$1,$-$1) & 
$-$\ts\frac{3}{2}+1=$-$\half \\ 
\vspace*{-0.35cm} &&& \\
u & (1,1,2,$-$2,$-$2) & (0,1,1,$-$1,$-$1) &
\half- 1=$-$\half \\
\vspace*{-0.35cm} &&& \\
\bar e^* & (1,1,4,$-$3,$-$3) & (0,1,1,$-$1,$-$1) &
\ts\frac{3}{2}-1=\half \\
\hline 
\end{array}
\end{eqnarray*}
\vspace*{-0.5cm}
\label{tab10}
\end{table}

\noindent
Within this table the decompositions of $Q$ 
into $Q_0$ and $Q_s$ are, respectively,
\bea
(0,1,1,\minus 1,\minus 1)&=&(1,1,1,\minus 2,\minus 1)+(\minus 1,0,0,1,0),\nn\\
(0,1,1,\minus 1,\minus 1)&=&(1,1,0,\minus 1,\minus 1)+(\minus 1,0,1,0,0),\nn\\
(0,1,1,\minus 1,\minus 1)&=&(1,1,0,\minus 1,\minus 1)+(\minus 1,0,1,0,0)\nn.
\eea

\section{Implications of a Theta Vacuum}
\label{sec:theta}

The last topic of concern is the effect of a theta vacuum 
\beq
\label{ltheta}
\calL_\theta = \frac{\theta g^2}{8\pi^2}\,\tr\,\bm E \cdot \bm B
\eeq
upon the SU(5) monopole and dyon spectrum. Although (\ref{ltheta}) is a total
divergence, and hence just a boundary term in the action, it does have a
physical effect. Witten showed that generically an electric charge is 
induced on a monopole through the asymptotic gauge potential~\cite{witten}.

A central feature of this theta vacuum is that it
violates parity and charge-parity maximally. This occurs because $\calL_\theta$
is both $\calP$ and $\calC\calP$ odd; whereas the usual Yang-Mills
Lagrangian is even. These violations appear
explicitly in the dyon spectra (\ref{ebtheta}) below.

In the context of a dual standard model this parity violation strongly suggests
that a theta vacuum should play a central role. Indeed, 
without including parity violation in a very unnatural manner, it is 
difficult to conceive of another method for incorporating parity
violation within the SU(5) dyon spectrum~\cite{vachdsm}. 

That a theta vacuum induces electric charge can be seen directly through
the interaction of a monopole with a gauge field $(\phi,\bm a)$. 
Following an argument of Coleman's~\cite{mon} 
the electric and magnetic fields  
\beq
\bm E = \bm\nabla \phi,\hspace{1em}
\bm B = \bm\nabla \wedge \bm a + \frac{1}{2g} \frac{\rhat}{r^2}M, 
\eeq
are substituted into (\ref{ltheta}) to give, upon integration by parts,
\beq
L_\theta = \int {\rm d}^3\bm r\  \calL_\theta =
 - \frac{\theta g}{2\pi} \int {\rm d}^3\bm r\  \de^3(\bm r)\,\tr\, \phi M.
\eeq
But this is precisely the interaction between the
gauge potential and an electric charge $Q_\theta = - \ts\frac{\theta}{2\pi}M$.
Therefore a theta vacuum induces electric charge.
This argument also carries through for a dyon $(Q,M)$, resulting in the 
electric and magnetic fields
\beq
\label{ebtheta}
\bm E \sim \frac{g}{4\pi} \frac{\rhat}{r^2} 
(Q - \frac{\theta}{2\pi}M),\hspace{3em}
\bm B \sim \frac{1}{2g}\frac{\rhat}{r^2}M.
\eeq

That parity and charge-parity are violated is explicit within the spectrum of 
dyons in (\ref{ebtheta}).
For instance the monopole/anti-monopole spectrum $(0,\pm M)$ becomes
$\pm(-\theta M/2\pi,M)$, which is not symmetric under either (\ref{p}) or 
(\ref{cp}). Also parity violation in the $(Q\mp\frac{\theta M}{2\pi},\pm M)$ 
dyon spectrum is seen in fig.~\ref{fig2} below.

\begin{figure}[ht]
\begin{center}
\epsfxsize=20em \epsfysize=10em \epsffile{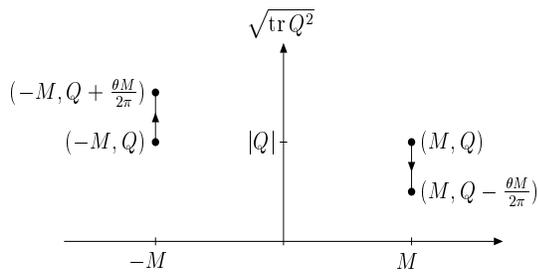}
\end{center}
\caption{$(Q,\pm M)$ dyons in a theta vacuum ({\em c.f.} fig.~\ref{fig1}).}
\label{fig2}
\end{figure}

Associated with the induced charge there will be an extra electric interaction
between electric charges and monopoles. An interesting property of
this interaction is that it is always associated with an Abelian gauge 
symmetry, even for non-Abelian generators $M$. Consequently
an electric charge $Q$ will interact with a magnetic monopole $M$ through
a Coulomb potential
\beq
V(r)\sim -\frac{\theta g^2}{2\pi}\frac{\tr\,{QM}}{4\pi r}.
\eeq
In addition there will also be the usual electric-magnetic interaction between
the charge and monopole.

This theta induced interaction is Abelian because of the global properties 
of charge around a monopole. Sec.~(\ref{secglob})
described how an electric charge can
only be consistently placed on a monopole if $[Q,M]=0$; otherwise there is 
an infinite energy string singularity. This leads to a group $H_A$ of 
globally allowed symmetries, which is relevant to charge-monopole interactions.
The point is that $M$ defines an Abelian symmetry $\U(1)_M$ such that
$H_A=\U(1)_M\x H'_A/\calZ$, which follows from $H_A$ centralising $M$. 
Consequently the induced theta interaction is always Abelian.

It is interesting that this induced theta interaction can provide a suitable 
binding potential $V(r)$ between some monopoles and electric 
charges, as discussed in sec.~(\ref{sec:spin}). Also notable is the spectrum 
of dyonic composites for which this potential is binding will maximally 
violate parity. 

\section{Conclusion}

This review has aimed to present a coherent picture of the monopoles
and dyons in SU(5) gauge unification. Their behaviour is fairly intricate,
although it does fit together in a coherent manner. A central point is
that these monopoles have properties which naturally apply
to the construction of a dual standard model. This supports Vachaspati's 
original proposal that the elementary particles might originate as monopoles 
from a magnetic gauge unification of the standard model.

The behaviour of these monopoles/dyons and their relation to a dual
standard model is as follows:

\noindent
(i) The central concept is that a charge may either be represented as sourcing
$A^0\sim e/4\pi r$ or as a monopole in the dual gauge potential 
$\tilde{A}_\varphi\sim e(1-\cos\theta)/4\pi$. Both descriptions are
expected to be equivalent by electric-magnetic duality. Therefore, in
principle, there is an alternative, dual, formulation of the standard model.
Such a formulation may uncover a hidden simplicity and regularity of form 
that underlies the conventional description.

\noindent
(ii) For scalar masses much less than gauge masses the magnetic charges
of the stable SU(5) monopoles are in one-to-one identification with the 
electric charges of one generation of elementary particles. This suggests 
that a dual formulation of the standard model should be constructed around 
the properties of SU(5) monopoles~\cite{vachdsm,vach95}.

\noindent
(iii) The gauge degeneracy of these five stable monopoles has an analogous
structure to the gauge degeneracy of each associated elementary 
particle~\cite{me:gauge}. This supports a duality in their gauge interactions.

\noindent
(iv) All stable SU(5) monopoles appear to have no intrinsic angular momentum. 
This means that some modification of their spectrum is required to make them 
spin and hence give a realistic dual standard model. 

\noindent
(v) A crucial point is that if dyons are considered instead of monopoles, 
then these may have intrinsic angular momentum. This strongly suggests
that a realistic dual standard model should be formulated around the 
properties of SU(5) dyons~\cite{vachdsm,vach95,vachspin}.

\noindent
(vi) The simplest dyons are scalar boson-monopole composites. 
Many of these have one-half angular momentum~\cite{vachspin}. 
However it is unclear how the dyons with one-half angular 
momentum are preferentially selected. It is also unclear how these dyonic 
composites are stabilised (although see (vii)).

\noindent
(vii) In addition to the dyons in (v) the SU(5) model also contains dyons
associated with a semi-classical quantisation of the electric field around a
monopole. Such dyons also appear to have one-half angular 
momentum~\cite{dixon84}, although there are many issues that have not 
been fully understood.

\noindent
(viii) A theta vacuum induces extra electric charge on each 
dyon/monopole~\cite{witten}. This may be important for
incorporating parity violation within a dual standard 
model~\cite{vachdsm,vach95}. It also has
relevance to the dyon spectrum because it induces a binding force that
stabilises some dyons.

The implications of a theta vacuum for constructing a realistic dual 
standard model are investigated from the two alternative viewpoints of
ref.~\cite{steer} and ref.~\cite{comp}.

\acknowledgments

Part of this work was supported by a junior research fellowship at
King's College, Cambridge. I thank H-M.~Chan, T.~Kibble, D.~Steer and
S.~Tsou for help and advice. I am also indebted to T.~Vachaspati for
his many helpful comments regarding this work.


\end{document}